# Resilient Operation of Transportation Networks via Variable Speed Limits

A. Yasin Yazıcıoğlu, Mardavij Roozbehani, and Munther A. Dahleh
Laboratory for Information and Decision Systems
Massachusetts Institute of Technology, Cambridge, MA 02139, USA
yasiny@mit.edu, mardavij@mit.edu, dahleh@mit.edu

*Abstract*— In this paper, we investigate the use of variable speed limits for resilient operation of transportation networks, which are modeled as dynamical flow networks under local routing decisions. In such systems, some external inflow is injected to the so-called origin nodes of the network. The total inflow arriving at each node is routed to its operational outgoing links based on their current particle densities. The density on each link has first order dynamics driven by the difference of its incoming and outgoing flows. A link irreversibly fails if it reaches its jam density. Such failures may propagate in the network and cause a systemic failure. We show that larger link capacities do not necessarily help in preventing systemic failures under local routing. Accordingly, we propose the use of variable speed limits to operate the links below their capacities, when necessary, to compensate for the lack of global information and coordination in routing decisions. Our main result shows that systemic failures under feasible external inflows can always be averted through a proper selection of speed limits if the routing decisions are sufficiently responsive to local congestion and the network is initially uncongested. This is an attractive feature as it is much easier in practice to adjust the speed limits than to build more physical capacity or to alter routing decisions that are determined by social behavior.

## I. INTRODUCTION

Resilience is a critical aspect in the design and operation of infrastructure systems such as transportation, power, water, and communication networks. In many applications, the network faces various disturbances and operates under the actions taken by users with limited information about the system. Such a distributed operation of the network often results in a suboptimal global performance (e.g., [1]), and it may even lead to cascading failures with severe systemic consequences (e.g., [2], [3]). Accordingly, there has been a significant research interest in modeling cascading failures on networks (e.g., [3], [4], [5]), investigating the influence of network topology on failure propagation (e.g., [6], [7]), and designing networks (e.g., [8], [9]) and control policies (e.g., [2], [3], [10], [11]) for resilient operation.

The main focus of this paper is on transporation networks under local routing decisions, which are modeled as dynamical flow networks as in [2], [3]. In such networks, some external inflow is injected to the so-called origin nodes. The total inflow arriving at each node is routed to its operational outgoing links based on their current particle densities. The density on each link has first order dynamics driven by the difference of its incoming and outgoing flows. A link irreversibly fails if it reaches its jam density. Such failures may propagate in the network since the outflow of a link drops to zero when there are no operational links in its immediate downstream. A systemic failure is observed if the long-term average inflow to the destination nodes (throughput) is less than the total external inflow.

Although links with larger capacities can sustain larger flows, the link capacities do not have a monotonic influence on the throughput of locally routed network flows [12]. A link with a larger capacity may attract a larger demand under local routing, and the resulting flow may cause a systemic failure by overloading some critical links in its downstream. Such failures can be avoided if the dynamics yield a sufficient back-propagation of congestion effects through the densities [10]. Otherwise, additional control and communication mechanisms are needed to mitigate the fragility arising from local decisions. One approach in this direction is to make the global network state accessible via communications. For instance, GPS-based route guidance systems are used in road traffic networks to provide the drivers with real-time traffic data. However, in distributed systems, the availability of more information does not guarantee the desired global performance in the absence of a properly coordinated response (e.g., [13], [14], [15]). Alternatively, some system objectives can be achieved via a controlled operation of the network, without requiring any global information and coordination in the user decisions (e.g., [12], [16], [17]).

In this paper, we investigate the provably correct use of variable speed limits as a flow control mechanism for avoiding systemic failures due to local routing decisions. Our work is closely related to [2], [3], [10] with the following main differences: We consider flows that are non-monotonic in densities due to congestion collapse, and we incorporate the speed limits as external control inputs in the flow functions. We show that systemic failures due to local routing decisions can always be averted through a proper selection of speed limits if the routing decisions are sufficiently responsive to local congestion and the initial densities on the links are sufficiently small. Accordingly, throughput optimality can be ensured without requiring any global information and coordination in the routing decisions, even when there is not a sufficient back-propagation of congestion effects. A tractable method of designing such speed limits under some limited knowledge of the routing behavior is also provided in the paper.

The organization of this paper is as follows: Section II provides some preliminaries. Section III presents the dynamical model of locally routed network flows. Section IV presents some motivation and examples. Section V presents our main results. Finally, Section VI concludes the paper.

## II. PRELIMINARIES

### A. Notation

For any finite set $A$ with cardinality $|A|$, we use $\mathbb{R}^A$ ($\mathbb{R}_+^A$, $\mathbb{R}_{++}^A$) to denote the space of real-valued (nonnegative-real-valued, positive-real-valued) $|A|$-dimensional vectors whose components are indexed by the elements of $A$. Accordingly, for any $a \in A$ and $x \in \mathbb{R}^A$, $x_a \in \mathbb{R}$ denotes the corresponding entry of $x$. Similarly, for any $A' \subseteq A$, we use $x_{A'} \in \mathbb{R}^{A'}$ to denote the $|A'|$-dimensional vector consisting only of the components of $x$ whose indices are in $A'$. For any pair of vectors $\underline{x}, \bar{x} \in \mathbb{R}^A$, we use $[\underline{x}, \bar{x})$ to denote the set of vectors $x \in \mathbb{R}^A$ such that $\underline{x} \leq x < \bar{x}$, i.e., $\underline{x}_a \leq x_a < \bar{x}_a$ for all $a \in A$. The all-zero vector, its size being clear from the context, will be denoted by $\mathbf{0}$.

### B. Graph basics

A directed graph, $\mathcal{G} = (\mathcal{V}, \mathcal{E})$, consists of a set of nodes, $\mathcal{V}$, and a set of edges, $\mathcal{E} \subseteq \mathcal{V} \times \mathcal{V}$, given by ordered pairs of nodes. A graph is a multi-graph if multiple edges are allowed between the nodes, i.e., if $\mathcal{E}$ is a multi-set. Each $(v, w) \in \mathcal{E}$ denotes a link from $v$ (the tail) to $w$ (the head). A path is a sequence of nodes such that $(v, w) \in \mathcal{E}$ for any pair of nodes $v, w \in \mathcal{V}$ that are consecutive in the sequence. For each $v \in \mathcal{V}$, we use $\mathcal{E}_v^-$ and $\mathcal{E}_v^+$ to denote the corresponding sets of incoming and outgoing links, respectively. Similarly, for any $\mathcal{U} \subseteq \mathcal{V}$,

$$\mathcal{E}_\mathcal{U}^- = \{(v, w) \in \mathcal{E} \mid v \notin \mathcal{U}, w \in \mathcal{U}\},$$

$$\mathcal{E}_\mathcal{U}^+ = \{(v, w) \in \mathcal{E} \mid v \in \mathcal{U}, w \notin \mathcal{U}\}.$$

## III. SYSTEM MODEL

In this section, we provide some definitions and the dynamical model of locally routed flows.

**Definition** (*Flow Network*): A flow network is a directed multi-graph $\mathcal{G} = (\mathcal{V}, \mathcal{E})$ such that there exists a path from any node to some destination node $v \in \mathcal{V}_D$, where

$$\mathcal{V}_D = \{v \in \mathcal{V} \mid \mathcal{E}_v^+ = \emptyset\}.$$

Furthermore, the set of origin nodes $\mathcal{V}_O$ is defined as

$$\mathcal{V}_O = \{v \in \mathcal{V} \mid \mathcal{E}_v^- = \emptyset\}.$$

The nodes that are neither an origin nor a destination constitute the set of intermediate nodes $\mathcal{V}_I = \mathcal{V} \setminus \mathcal{V}_O \setminus \mathcal{V}_D$.

In flow networks, each link $e \in \mathcal{E}$ has a flow $f_e \in \mathbb{R}_+$ that is related to the average number of particles per unit length (density) $\rho_e \in \mathbb{R}_+$ and their average speed $s_e \in \mathbb{R}_+$ as

$$f_e = \rho_e s_e. \tag{1}$$

In this setting, the particles are assumed to move at the maximum admissible speed, i.e.,

$$s_e = \min(\bar{s}_e(\rho_e), u_e), \tag{2}$$

where $\bar{s}_e(\rho_e)$ denotes the maximum speed at which the particles can travel along $e$ when the density is $\rho_e$, and $u_e \in [0, \bar{u}_e]$ is the speed limit imposed on the link. In transportation networks, $\bar{s}_e(\rho_e)$ is a monotonically decreasing function since a higher density corresponds to a shorter following distance, for which the maximum safe speed is lower. Furthermore, each $\bar{u}_e \in \mathbb{R}_{++}$ is typically set based on some safety, efficiency, and environmental considerations.

In light of (1) and (2), the flow on each link is a function of its density and speed limit, i.e.,

$$f_e : [0, \bar{\rho}_e] \times [0, \bar{u}_e] \mapsto [0, \bar{f}_e],$$

where $\bar{f}_e \in \mathbb{R}_{++}$ denotes the finite capacity of the link, that is, the maximum flow that can be sustained by the link, and $\bar{\rho}_e \in \mathbb{R}_{++}$ is the finite jam density, for which $\bar{s}_e(\bar{\rho}_e) = 0$. Accordingly,

$$f_e(0, u_e) = f_e(\bar{\rho}_e, u_e) = 0, \quad \forall u_e \in [0, \bar{u}_e]. \tag{3}$$

Note that the flow becomes a function of only the density when the speed limit is constant. For instance, $f_e(\rho_e, \bar{u}_e)$ represents how the flow depends on the density when the link is operated constantly with the maximum speed limit.

**Assumption 1** Each $f_e(\rho_e, \bar{u}_e)$ is continuous in $\rho_e$, and there exists a congestion threshold $\rho_e^c \in (0, \bar{\rho}_e)$ such that $f_e(\rho_e, \bar{u}_e)$ is strictly increasing on the interval $(0, \rho_e^c)$ and strictly decreasing on the interval $(\rho_e^c, \bar{\rho}_e)$.

The structure in Assumption 1 is typical for the flow functions in transportation networks (e.g., [18]). Accordingly, the capacity of link $e$ is $\bar{f}_e = f_e(\rho_e^c, \bar{u}_e)$, and the link is said to be congested when $\rho_e > \rho_e^c$.

We focus on networks where the total inflow to each node is routed locally to its operational outgoing links based on their current densities. Such routing behavior subject to the conservation of flow is modeled via local routing policies.

**Definition** (*Local Routing Policy*): For any flow network $\mathcal{G} = (\mathcal{V}, \mathcal{E})$ with the finite jam densities $\bar{\rho} \in \mathbb{R}_{++}^{\mathcal{E}}$, a local routing policy $\mathcal{R}$ is a family of functions

$$\mathcal{R}^v : [\mathbf{0}, \bar{\rho}_{\mathcal{E}_v^+}] \times \mathbb{R}_+ \mapsto \mathbb{R}_+^{\mathcal{E}_v^+}, \ v \in \mathcal{V} \setminus \mathcal{V}_D,$$

such that, for any inflow $\mu_v \in \mathbb{R}_+$,

$$\sum_{e \in \mathcal{E}_v^+} \mathcal{R}_e^v(\rho_{\mathcal{E}_v^+}, \mu_v) = \mu_v, \ \forall \rho_{\mathcal{E}_v^+} \neq \bar{\rho}_{\mathcal{E}_v^+}, \tag{4}$$

$$\mathcal{R}_e^v(\rho_{\mathcal{E}_v^+}, \mu_v) = 0, \ \forall e \in \mathcal{E}_v^+ : \rho_e = \bar{\rho}_e, \tag{5}$$

where $\mathcal{R}_e^v(\rho_{\mathcal{E}_v^+}, \mu_v)$ is the flow routed to the link $e \in \mathcal{E}_v^+$.

**Definition** (*Locally Routed Flow*): For any flow network $\mathcal{G} = (\mathcal{V}, \mathcal{E})$, flow functions $f$, local routing policy $\mathcal{R}$, and external inflow $\lambda \in \mathbb{R}_+^{\mathcal{V}_O}$, the locally routed flow $(\mathcal{G}, f, \mathcal{R}, \lambda)$ is a dynamical system such that, for every $e = (v, w) \in \mathcal{E}$,

$$\dot{\rho}_e = \begin{cases} \mathcal{R}_e^v(\rho_{\mathcal{E}_v^+}, \mu_v) - f_e(\rho_e, u_e), & \text{if } \rho_{\mathcal{E}_w^+} \neq \bar{\rho}_{\mathcal{E}_w^+}, \\ \mathcal{R}_e^v(\rho_{\mathcal{E}_v^+}, \mu_v), & \text{o.w.} \end{cases} \quad (6)$$

where $\mu_v$ denotes the total inflow to $v$, i.e.,

$$\mu_v = \begin{cases} \lambda_v, & \text{if } v \in \mathcal{V}_O, \\ \sum_{j \in \mathcal{E}_v^-} f_j(\rho_j, u_j), & \text{o.w.} \end{cases} \quad (7)$$

and $u_e \in [0, \bar{u}_e]$ is the speed limit imposed on $e$.

In light of (3) and (5), each $\bar{\rho}_e$ is an equilibrium point under (6). As such, a link irreversibly stops carrying flow (fails) if it reaches its jam density. Furthermore, since (6) implies $\dot{\rho}_e \geq 0$ for any $e = (v, w) \in \mathcal{E}$ such that $\rho_{\mathcal{E}_w^+} = \bar{\rho}_{\mathcal{E}_w^+}$, link failures can propagate in the network. While some failures may not have a severe impact on the overall performance, others can cause a cascading failure of some critical links and lead to a non-transferring system.

**Definition** (*Transferring System*): A locally routed flow is transferring if the long-term average inflow to the destination nodes (throughput) is equal to the total external inflow, i.e.,

$$\liminf_{t \to \infty} \frac{1}{t} \int_0^t \sum_{v \in \mathcal{V}_D} \mu_v(\tau) d\tau = \sum_{v \in \mathcal{V}_O} \lambda_v.$$

In transferring systems, the amount of particles that are unable to reach a destination node remains bounded as time goes to infinity. In contrast, non-transferring systems experience a systemic failure, i.e., at least one origin node, which has a non-zero external inflow, becomes disconnected from the network due to link failures (see Lemma 5.1). As such, eventually it becomes impossible to send the corresponding portion of the external inflow through the network.

## IV. MOTIVATION

Consider any locally routed network flow $(\mathcal{G}, f, \mathcal{R}, \lambda)$, where all the links are operated constantly under their respective maximum admissible speed limits, i.e., $u(t) = \bar{u}$, so that they all can sustain flows up to their respective capacities. If such a system is non-transferring despite the initial densities being sufficiently small (e.g., $\rho(0) = \mathbf{0}$), this failure is due to one of the two possible reasons:

- The external inflow $\lambda$ is too large (infeasible) and cannot be transferred under any routing policy.
- The external inflow $\lambda$ is feasible, yet the routing decisions lead to failure.

In this regard, an external inflow is feasible if the corresponding set of feasible equilibrium flows, i.e.,

$$\mathcal{F}(\bar{f}, \lambda) = \{ f \in \mathbb{R}_+^{\mathcal{E}} \mid f_e \leq \bar{f}_e, \forall e \in \mathcal{E} \\ \lambda_v = \sum_{e \in \mathcal{E}_v^+} f_e, \forall v \in \mathcal{V}_O \\ \sum_{e \in \mathcal{E}_v^-} f_e = \sum_{e \in \mathcal{E}_v^+} f_e, \forall v \in \mathcal{V}_I \} \quad (8)$$

is nonempty. The feasibility of an external inflow can be checked based on the max-flow min-cut theorem [19], which implies that $\mathcal{F}(\bar{f}, \lambda) \neq \emptyset$ if and only if

$$\sum_{e \in \mathcal{E}_{\mathcal{U}}^+} \bar{f}_e - \sum_{v \in \mathcal{U}} \lambda_v \geq 0, \ \forall \mathcal{U} \subseteq \mathcal{V} \setminus \mathcal{V}_D. \quad (9)$$

Accordingly, for any infeasible external inflow, there exists some $\mathcal{U} \subseteq \mathcal{V} \setminus \mathcal{V}_D$, which receives a total external inflow larger than the total maximum flow sustainable by its outgoing links. In that case, the total density on the links within $\mathcal{U}$ keeps increasing until all the outgoing links of some origin node in $\mathcal{U}$ fail. Hence, infeasible external inflows lead to a non-transferring system, regardless of the routing behavior.

In light of (9), systemic failures due to infeasible external inflows cannot be avoided without reducing the external inflow or building more capacity in the network. However, this is not the case for failures due to routing decisions. In fact, we will show that such failures can be avoided by using the speed limits to intentionally operate some of the links below their capacity, provided that the routing policy is sufficiently congestion aware.

**Definition** (*Congestion Awareness*): For any flow network $\mathcal{G} = (\mathcal{V}, \mathcal{E})$ and flow functions $f$, a local routing policy $\mathcal{R}$ is congestion aware at a density profile $\rho^* \in [\mathbf{0}, \bar{\rho}]$ if, for every $e = (v, w) \in \mathcal{E}$,

$$\rho_e \geq \rho_e^* \Rightarrow \mathcal{R}_e^v(\rho_{\mathcal{E}_v^+}, \mu_v) \leq \phi_e(\rho_e), \ \forall \mu_v \leq \sum_{j \in \mathcal{E}_v^+} \phi_j(\rho_j), \quad (10)$$

where $\phi_e(\rho_e)$ denotes the maximum sustainable inflow, i.e.,

$$\phi_e(\rho_e) = \begin{cases} \bar{f}_e, & \text{if } \rho_e \leq \rho_e^c, \\ f_e(\rho_e, \bar{u}_e), & \text{o.w.} \end{cases} \quad (11)$$

Note that, for any $\rho_e \in [0, \bar{\rho}_e)$, $\phi_e(\rho_e)$ is the maximum constant inflow that would not drive the link to failure when its current density is $\rho_e$. As such, (10) indicates some local failure-avoidance behavior. In particular, when the density of a link $e = (v, w)$ exceeds the corresponding $\rho_e^*$, the flow routed to $e$ does not exceed its maximum sustainable inflow as long as $v$ is receiving a sufficiently small $\mu_v$.

### A. Motivating Example

Consider the flow network in Fig. 1a. Let all the links have similar characteristics except for their cross-sectional dimensions. Accordingly, let $f_0(\rho_0, u_0)$ denote a normalized flow function such that

$$f_e(\rho_e, u_e) = c_e f_0\left(\frac{\rho_e}{c_e}, u_e\right), \quad \forall e \in \mathcal{E},$$

where $c_e \in \mathbb{R}_{++}$ is determined by the cross-sectional dimension of the link $e$. For instance, each $c_e$ may correspond to the number of lanes in $e$, whereas $f_0$ denotes the flow function for a single lane. Let all the links be constantly operated under the maximum speed limit $\bar{u}_0 \in \mathbb{R}_{++}$, and let $f_0(\rho_0, \bar{u}_0)$ be as illustrated in Fig. 1b. Accordingly, the capacity of each link is $\bar{f}_e = c_e \bar{f}_0$ and provided next to itself in Fig. 1a. Suppose that initially there are no particles in the network, i.e., $\rho(0) = \mathbf{0}$, and an external inflow of $\lambda = 6\bar{f}_0$ is injected. Note that this is a feasible external inflow for this network since the minimum value for the left side of (9) is $\bar{f}_0$, which is obtained for $\mathcal{U} = \{1, 2\}$.

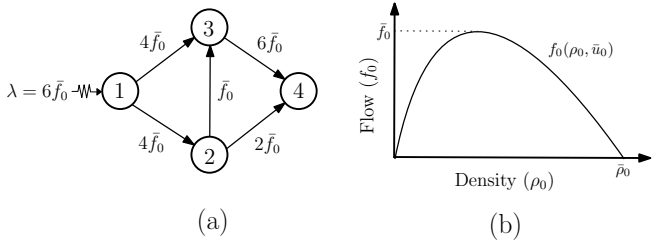

Fig. 1. A flow network along with the external inflow and link capacities are illustrated in (a), and the normalized flow function is shown in (b).

Since the external inflow is feasible and $\rho(0) = \mathbf{0}$, whether the system is transferring depends on the routing policy. Consider a local routing policy $\mathcal{R}$ such that the flow routed to each edge is proportional to its maximum sustainable inflow, that is, for every $e = (v, w) \in \mathcal{E}$,

$$\mathcal{R}_e^v(\rho_{\mathcal{E}_v^+}, \mu_v) = \mu_v \frac{\phi_e(\rho_e)}{\sum_{j \in \mathcal{E}_v^+} \phi_j(\rho_j)}, \forall \rho_{\mathcal{E}_v^+} \neq \bar{\rho}_{\mathcal{E}_v^+}. \quad (12)$$

In light of (10), this routing policy is congestion aware at any $\rho^* \in [\mathbf{0}, \bar{\rho}]$. Furthermore, it can be shown that when $\rho(0) \in [\mathbf{0}, \rho^c]$ and the maximum speed limits are constantly applied as $u(t) = \bar{u}$, the tail of the first link that exceeds its congestion threshold under this policy must have received an inflow larger than the total capacity of its outgoing links. For the network in Fig. 1a, it can be easily verified that node 1 (since $\lambda$ is less than its total outgoing capacity) and node 3 (since its total incoming capacity is less than its total outgoing capacity) never receive such excessive inflows. That is also true for node 2 since $\mathcal{R}^1$ routes $\lambda$ evenly to the links in $\mathcal{E}_1^+$ as long as they are both uncongested. Accordingly, $\mu_2$ monotonically increases to $3\bar{f}_0$, which is equal to the total capacity of the links in $\mathcal{E}_2^+$. Hence, none of the links exceed their congestion thresholds in this example, and the resulting system $(\mathcal{G}, f, \mathcal{R}, \lambda)$ is transferring.

As a second example, consider the same network when $c_{(2,4)}$ is reduced by one due to some disturbance. For instance, such a reduction can be due to one of the lanes in the link $(2, 4)$ being closed due to maintenance or an accident. Note that the external inflow of $\lambda = 6\bar{f}_0$ is still feasible under the resulting link capacities as shown in Fig. 2a. Same as the previous case, consider the routing policy $\mathcal{R}$ that satisfies (12) for the flow functions in this example, and let the system have $\rho(0) = \mathbf{0}$ and $u(t) = \bar{u}$. Under the resulting dynamics, the densities of the links in $\mathcal{E}_1^+$ initially follow the same trajectory as before. However, once $\mu_2$ exceeds $2\bar{f}_0$ (the total capacity of the links in $\mathcal{E}_2^+$), the links in $\mathcal{E}_2^+$ will be driven to their jam densities. Note that $\lambda$ is not feasible for the remaining network once those two links fail, and the failures in Fig. 2d are inevitable. As such, $(\mathcal{G}, f, \mathcal{R}, \lambda)$ is non-transferring for this example.

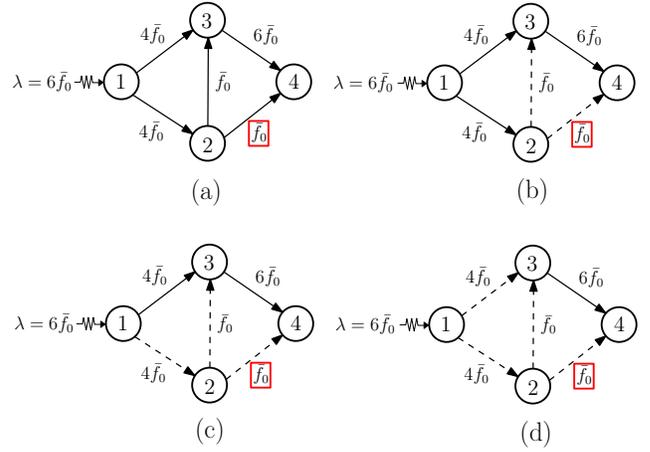

Fig. 2. A reduction in the capacity of the link $(2, 4)$ as shown in (a) leads to a cascade of failures as illustrated in (b)-(c)-(d).

To avoid the failures in the second example, a larger portion of $\lambda$ should be routed to the link $(1, 3)$ before the links in $\mathcal{E}_2^+$ are overloaded. However, local routing policies only respond to the densities on the immediate links. Hence, even the routing policies such as (12), which are congestion aware at every density profile in $[\mathbf{0}, \bar{\rho}]$, may fall short in circumventing systemic failures when the links can sustain flows that may overload some critical links further in their downstream. In this regard, the large capacity of the link $(1, 2)$ was actually a liability in the second example. As we will show later in the paper, the systemic failure in this example could actually be avoided by reducing the capacity of the link $(1, 2)$ so that its outflow never exceeds $2\bar{f}_0$.

### B. Capacity Allocation via Speed Limits

As illustrated in the previous section, having large link capacities may also lead to systemic failures when the routing decisions are local. In this section, we present how speed limits can be used to address this issue by intentionally operating certain links below their capacities when necessary.

Given a flow function $f_e(\rho_e, \bar{u}_e)$ and some $\bar{f}_e^* \in [0, \bar{f}_e]$, suppose that we would like to design the speed limit $u_e$ such that the flow on the link never exceeds $\bar{f}_e^*$. In order to avoid unnecessary delays, it would also be desired to keep $u_e$ as large as possible. The solution to this problem depends on the available flexibility in the design. For instance, if a constant speed limit is required, then (1) and (2) together with Assumption 1 imply that the solution is

$$u_e = \frac{\bar{f}_e^*}{\hat{\rho}_e}, \quad (13)$$

where

$$\hat{\rho}_e = \max\{\rho_e \in [0, \bar{\rho}_e] \mid f_e(\rho_e, \bar{u}_e) = \bar{f}_e^*\}. \quad (14)$$

Such a constant speed limit may be too conservative when the density on the link is low. Alternatively, a feedback law can induce the desired capacity more efficiently. For instance,

the following policy achieves this task while allowing for the maximum possible speed at all times:

$$u_e(\rho_e) = \begin{cases} \bar{u}_e, & \text{if } f_e(\rho_e, \bar{u}_e) \leq \bar{f}_e^*, \\ \frac{\bar{f}_e^*}{\rho_e}, & \text{o.w.} \end{cases} \quad (15)$$

Fig. 3 illustrates how the speed limits in (13) and (15) influence the flow on a link for a typical flow function. Accordingly, if $\bar{f}_e^*$ is chosen properly, such an operation of the link ensures that the link accumulates a higher density itself rather than overloading some critical links further in its downstream. This way, potential failures due to the lack of global information can be avoided if the local routing decisions properly respond to the increasing density on the link. In the next section, we provide our formal results regarding such use of speed limits to avert systemic failures in locally routed flows.

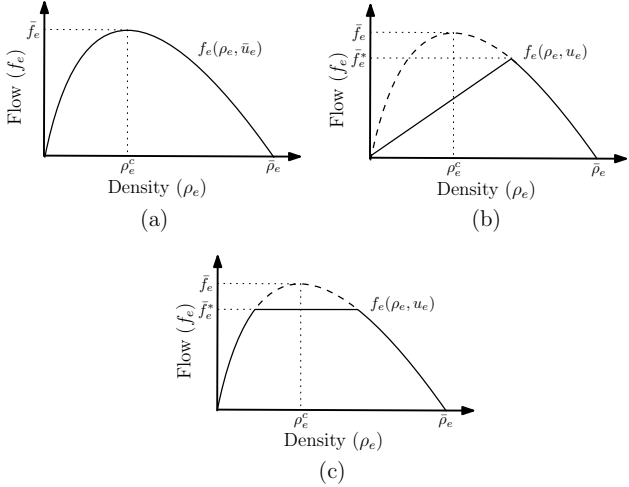

Fig. 3. Capacity allocation via the speed limit policies in (13) and (15) are illustrated in (b) and (c), respectively. The solid curves show the resulting flow functions, the dashed curves illustrate the deviation from $f_e(\rho_e, \bar{u}_e)$.

## V. MAIN RESULTS

Our main result (Theorem 5.2) shows that, for any locally routed flow, if the routing policy is congestion aware at a density profile $\rho^* \in [\mathbf{0}, \bar{\rho}]$ that is compatible with the external inflow, then there exist speed limits $u(t) \in [\mathbf{0}, \bar{u}]$ that ensure a transferring system for any initial condition $\rho(0) \in [\mathbf{0}, \rho^*]$. We start our derivations with Lemma 5.1, which will be used in proving Theorem 5.2.

**Lemma 5.1** *If a locally routed flow $(\mathcal{G}, f, \mathcal{R}, \lambda)$ is non-transferring, then there exists an origin node $v \in \mathcal{V}_O$ such that its external inflow is non-zero and all its outgoing links eventually reach their jam densities, i.e.,*

$$\lambda_v > 0, \quad \lim_{t \to \infty} \rho_{\mathcal{E}_v^+}(t) = \bar{\rho}_{\mathcal{E}_v^+}. \quad (16)$$

*Proof:* Since every local routing policy $\mathcal{R}$ satisfies (4) and (5), under the dynamics in (6) and (7),

$$\sum_{e \in \mathcal{E}} \dot{\rho}_e = \sum_{v \in \mathcal{V}_O : \rho_{\mathcal{E}_v^+} \neq \bar{\rho}_{\mathcal{E}_v^+}} \lambda_v - \sum_{v \in \mathcal{V}_D} \mu_v. \quad (17)$$

For the sake of contradiction, suppose that $(\mathcal{G}, f, \mathcal{R}, \lambda)$ is non-transferring, i.e.,

$$\liminf_{t \to \infty} \frac{1}{t} \int_0^t \sum_{v \in \mathcal{V}_D} \mu_v(\tau) d\tau < \sum_{v \in \mathcal{V}_O} \lambda_v, \quad (18)$$

and there is no $v \in \mathcal{V}_O$ satisfying (16). In that case, (17) and (18) together imply

$$\liminf_{t \to \infty} \frac{1}{t} \int_0^t \sum_{e \in \mathcal{E}} \dot{\rho}_e(\tau) d\tau > 0,$$

which leads to

$$\liminf_{t \to \infty} \sum_{e \in \mathcal{E}} \rho_e(t) = +\infty. \quad (19)$$

Note that (19) contradicts with every $\bar{\rho}_e$ being finite. Consequently, if $(\mathcal{G}, f, \mathcal{R}, \lambda)$ is non-transferring, then there exists $v \in \mathcal{V}_O$ satisfying (16). ∎

**Theorem 5.2** *For any locally routed flow $(\mathcal{G}, f, \mathcal{R}, \lambda)$, if $\mathcal{R}$ is congestion aware at some $\rho^* \in [\mathbf{0}, \bar{\rho}]$ such that the corresponding maximum sustainable inflows in (11) satisfy*

$$\sum_{e \in \mathcal{E}_{\mathcal{U}}^+} \phi_e(\rho_e^*) - \sum_{v \in \mathcal{U}} \lambda_v \geq 0, \ \forall \mathcal{U} \subseteq \mathcal{V} \setminus \mathcal{V}_D, \quad (20)$$

*then there exist speed limits $u(t) \in [\mathbf{0}, \bar{u}]$ such that $(\mathcal{G}, f, \mathcal{R}, \lambda)$ is transferring for any initial condition $\rho(0) \in [\mathbf{0}, \rho^*]$.*

*Proof:* Let $(\mathcal{G}, f, \mathcal{R}, \lambda)$ be any locally routed flow such that $\mathcal{R}$ is congestion aware at some $\rho^* \in [\mathbf{0}, \bar{\rho}]$ satisfying (20). Let $\bar{\mathcal{F}}^*(\phi(\rho^*), \lambda)$ be the set of feasible capacity allocations that are upper bounded by $\phi(\rho^*)$ and ensures that the total outgoing capacity is greater than the external inflow for each origin node and greater than the total incoming capacity for each intermediate node, i.e.,

$$\bar{\mathcal{F}}^*(\phi(\rho^*), \lambda) = \{\bar{f}^* \in \mathbb{R}_+^{\mathcal{E}} \mid \bar{f}_e^* \leq \phi_e(\rho_e^*), \ \forall e \in \mathcal{E}$$
$$\lambda_v \leq \sum_{e \in \mathcal{E}_v^+} \bar{f}_e^*, \ \forall v \in \mathcal{V}_O$$
$$\sum_{e \in \mathcal{E}_v^-} \bar{f}_e^* \leq \sum_{e \in \mathcal{E}_v^+} \bar{f}_e^*, \ \forall v \in \mathcal{V}_I \} \quad (21)$$

Based on (8), $\mathcal{F}(\phi(\rho^*), \lambda) \subseteq \bar{\mathcal{F}}^*(\phi(\rho^*), \lambda)$. Furthermore, due to (20), $\mathcal{F}(\phi(\rho^*), \lambda) \neq \emptyset$. Hence, $\bar{\mathcal{F}}^*(\phi(\rho^*), \lambda) \neq \emptyset$. Let the speed limits $u(t) \in [\mathbf{0}, \bar{u}]$ be assigned in accordance with (13) or (15) for any $\bar{f}^* \in \bar{\mathcal{F}}^*(\phi(\rho^*), \lambda)$. Such speed limits ensure that the flow on each link $e$ is upper bounded by $\bar{f}_e^*$. Accordingly, (7) and (21) together imply

$$\mu_v(t) \leq \sum_{e \in \mathcal{E}_v^+} \bar{f}_e^*, \ \forall v \in \mathcal{V}, \ \forall t \geq 0. \quad (22)$$

Let $\hat{\rho}$ denote the vector of densities such that each $\hat{\rho}_e$ satisfies (14). Due to Assumption 1, $\hat{\rho} \in [\rho^*, \bar{\rho}]$ and

$$\phi_e(\rho_e) \geq \phi_e(\hat{\rho}_e) = \bar{f}_e^*, \ \forall \rho_e \in [0, \hat{\rho}_e], \ \forall e \in \mathcal{E}. \quad (23)$$

As such, (10), (22) and (23) together imply that, under (6),

$$\rho_e = \hat{\rho}_e \Rightarrow \dot{\rho}_e \leq 0, \ \forall \rho \in [\mathbf{0}, \hat{\rho}], \ \forall e \in \mathcal{E}.$$

Hence, $[\mathbf{0}, \hat{\rho}]$ is positively invariant. Also, due to (21),

$$\sum_{e \in \mathcal{E}_v^+} \bar{f}_e^* > 0, \ \forall v \in \mathcal{V}_O : \lambda_v > 0,$$

which implies

$$\hat{\rho}_{\mathcal{E}_v^+} \neq \bar{\rho}_{\mathcal{E}_v^+}, \ \forall v \in \mathcal{V}_O : \lambda_v > 0. \quad (24)$$

Since $[\mathbf{0}, \hat{\rho}]$ is positively invariant, Lemma 5.1 and (24) together imply that $(\mathcal{G}, f, \mathcal{R}, \lambda)$ is transferring for any $\rho(0) \in [\mathbf{0}, \hat{\rho}]$. Since $\hat{\rho} \geq \rho^*$, we conclude that $(\mathcal{G}, f, \mathcal{R}, \lambda)$ is transferring for any $\rho(0) \in [\mathbf{0}, \rho^*]$. ∎

Theorem 5.2 implies that the systemic failures under any feasible external inflow can be averted through a proper selection of speed limits if the local routing policy is sufficiently congestion aware and the initial densities on the links are sufficiently small. For instance, since $\phi(\rho^c) = \bar{f}$ and any feasible external inflow satisfies (9), the theorem implies that for any $(\mathcal{G}, f, \mathcal{R}, \lambda)$ such that $\lambda$ is feasible and $\mathcal{R}$ is congestion aware at the congestion threshold $\rho^c$, there exist speed limits $u(t) \in [\mathbf{0}, \bar{u}]$ such that $(\mathcal{G}, f, \mathcal{R}, \lambda)$ is transferring for any $\rho(0) \in [\mathbf{0}, \rho^c]$. Accordingly, a proper choice of speed limits ensures optimal throughput by compensating for the lack of global information in routing decisions. We will conclude this section by providing a tractable method of generating such speed limits.

*A. Minimal Capacity Reduction with Maximal Speed Limits*

As shown in the constructive proof of Theorem 5.2, one way to use the speed limits for ensuring a transferring system is by inducing some capacities $\bar{f}^* \in \bar{\mathcal{F}}^*(\phi(\rho^*), \lambda)$. Since each $f_e(\rho_e, u_e)$ is monotonically increasing in $u_e$, inducing smaller capacities requires imposing lower speed limits. Hence, to employ this approach without introducing unnecessary delays on the individual links, it would be desired to realize some maximal $\bar{f}^*$. For instance, such a capacity allocation problem can be formulated as

$$\bar{f}^* = \underset{\bar{f}' \in \bar{\mathcal{F}}^*(\phi(\rho^*), \lambda)}{\arg\max} \alpha^\top \bar{f}', \quad (25)$$

where each $\alpha_e \in \mathbb{R}_{++}$ denotes the relative importance of having a larger capacity on $e$. Note that (25) is a linear program since $\bar{\mathcal{F}}^*(\phi(\rho^*), \lambda)$ is defined by the linear inequalities in (21). Since $\bar{\mathcal{F}}^*(\phi(\rho^*), \lambda)$ is non-empty (see the proof of Theorem 5.2) and bounded ($\bar{\mathcal{F}}^*(\phi(\rho^*), \lambda) \subseteq [\mathbf{0}, \phi(\rho^*)]$), a solution to (25) is guaranteed to exist. Once the capacity allocation problem is solved, a speed limit policy such as (15) can be used to realize $\bar{f}^*$ without unnecessary slowdowns.

As an application of this method, let us revisit the second scenario in Section IV-A. Given the external inflow and link capacities in Fig. 4a, suppose that we would like to assign the speed limits to ensure a transferring system for any local routing policy that is congestion aware at $\rho^c$ and any initial condition $\rho(0) \in [\mathbf{0}, \rho^c]$. For this example, (25) has a unique solution for $\rho^* = \rho^c$ and any $\alpha \in \mathbb{R}_{++}^{\mathcal{E}}$, i.e.,

$$\bar{f}_{(1,2)}^* = 2\bar{f}_0, \ \ \bar{f}_e^* = \bar{f}_e, \forall e \in \mathcal{E} \setminus \{(1,2)\},$$

which can be realized by using the maximal speed limits in (15) to achieve the desired performance guarantee. As such, the resulting speed limits would prevent the systemic failure illustrated in Fig. 2.

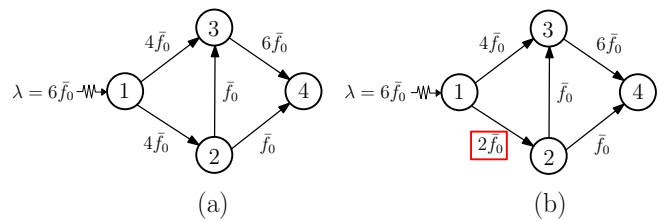

Fig. 4. For the network in (a), the solution to the minimal capacity reduction problem in (25) for $\rho^* = \rho^c$ and any $\alpha \in \mathbb{R}_{++}^{\mathcal{E}}$ is illustrated in (b). Accordingly, the systemic failure illustrated in Fig. 2 can be prevented by using variable speed limits to induce a lower capacity on the link $(1,2)$.

VI. CONCLUSIONS

In this paper, we investigated the use of variable speed limits for avoiding systemic failures in transportation networks, which were modeled as dynamical flow networks under local routing decisions. We showed that operating the links constantly at their full capacities may lead to systemic failures in such systems due to the lack of global information in routing decisions. In particular, the capacity of a link becomes a liability when it sustains a large flow that will overload some critical links further in its downstream. Accordingly, variable speed limits can be employed to operate the links below their full capacities when necessary to compensate for the lack of global information. Our main result shows that systemic failures under feasible external inflows can always be averted through a proper selection of speed limits if the routing decisions are sufficiently responsive to local congestion and the initial densities on the links are sufficiently small. A tractable method for designing such speed limit policies was also provided in the paper.

The results of this paper pave the way for the design of provably correct speed limit policies which not only ensure a transferring system but also optimize some additional performance measures such as the average time the particles spend in the network (average delay). As a future direction, we plan to investigate the design of such policies under limited knowledge of the routing behavior. Furthermore, we plan to investigate how the availability of more information (e.g., multi-hop densities) in the routing policies would affect the overall performance under such network operation.